\documentclass[a4paper,conference,compsoc]{IEEEtran}
\usepackage[top=0.75in, bottom=1.694in, left=0.514in, right=0.514in]{geometry}
\usepackage{amsmath}
\usepackage{amssymb} 
\usepackage{color}
\usepackage{graphicx}
\usepackage{subfigure}

\newtheorem{remark}{Remark}[section]

\ifCLASSOPTIONcompsoc
  \usepackage[nocompress]{cite}
\else
  \usepackage{cite}
\fi
\ifCLASSINFOpdf
\else
\fi
\usepackage[english]{babel}
\usepackage[latin1]{inputenc}
\usepackage[T1]{fontenc}

\begin{document}
%
\title{Energy saving for building heating \\ via a simple and efficient model-free control design: \\ First steps with computer simulations}


\author{\IEEEauthorblockN{Hassane Aboua\"{\i}ssa\IEEEauthorrefmark{1},
Ola Alhaj Hasan\IEEEauthorrefmark{2}, C\'{e}dric Join\IEEEauthorrefmark{3}\IEEEauthorrefmark{5}\IEEEauthorrefmark{6}, Michel Fliess\IEEEauthorrefmark{4}\IEEEauthorrefmark{5}, and Didier Defer\IEEEauthorrefmark{2}
 }
\IEEEauthorblockA{\IEEEauthorrefmark{1}LGI2A (EA 3926), Universit\'{e} d'Artois, 64200 B\'{e}thune, France. Email: hassane.abouaissa@univ-artois.fr}
\IEEEauthorblockA{\IEEEauthorrefmark{2}LGCgE (EA 4515), Universit\'{e} d'Artois, 64200 B\'{e}thune, France. \\Email: eng.3ola@gmail.com, didier.defer@univ-artois.fr}
\IEEEauthorblockA{\IEEEauthorrefmark{3}CRAN (CNRS, UMR 7039), Universit\'{e} de Lorraine, BP 239, 54506 Vand{\oe}uvre-l\`{e}s-Nancy, France. \\ Email: cedric.join@univ-lorraine.fr}
\IEEEauthorblockA{\IEEEauthorrefmark{4}LIX (CNRS, UMR 7161), \'Ecole polytechnique, 91128 Palaiseau, France. Email: Michel.Fliess@polytechnique.edu}
\IEEEauthorblockA{\IEEEauthorrefmark{5}AL.I.E.N., 7 rue Maurice Barr\`{e}s, 54330 V\'{e}zelise, France. Email: \{michel.fliess, cedric.join\}@alien-sas.com}
\IEEEauthorblockA{\IEEEauthorrefmark{6}Projet Non-A, INRIA Lille -- Nord-Europe, France.}}



%


\maketitle

\begin{abstract}
The model-based control of building heating systems for energy saving encounters severe physical, mathematical and calibration difficulties in the numerous attempts that has been published until now. This topic is addressed here via a new \emph{model-free} control setting, where the need of any mathematical description disappears. Several convincing computer simulations are presented. Comparisons with classic PI controllers and flatness-based predictive control are provided.   

\end{abstract}


%
\IEEEpeerreviewmaketitle

\section{Introduction}
The growing importance of energy saving explains the key r\^ole of heating, ventilation and air conditioning (HVAC). It raises many exciting questions (see, \textit{e.g.}, \cite{ka}). We concentrate here on the building heating system. This topic, according to \cite{Sam10}, has not attracted enough investigation from the control community. Let us mention here Boolean control, predictive control, PIDs, optimal control, nonlinear control, partial differential equations, flatness-based control, fuzzy systems, neural nets, \dots \ See, \textit{e.g.}, \cite{afram1,afram2,Ola_thesis,Ola16,Ola14,asw,balan,mines,bianchini,neural,haines,Hazyuk2012388,pde,koch,li,ma,Moac,Murray201498,Ooka20091538,PEG10,Sam10,siro,tur,wang}, \dots \
The more ``advanced'' control approaches, which are listed above, are model-based. Here again, like in most concrete situations, writing a ``good'' mathematical model, where constraints and perturbations might be severe, is quite beyond our reach especially if online calibration ought to be performed. Those facts explain why in industrial practice classic PIDs and Boolean control, which to a large extent preclude any mathematical modeling, are most popular in spite of some shortcomings:  poor performances and delicate tuning. 

The topic is addressed here via \emph{model-free control} and their corresponding \emph{intelligent} controllers \cite{ijc13}. This setting, 
\begin{itemize}
\item where the need of any precise mathematical description disappears, 
\item which is 
\begin{itemize}
\item inherently robust, since the perturbations are easily taken into account,
\item easy to implement both from software and hardware viewpoints,
\end{itemize}
\end{itemize}
has already been successfully applied a number of times, and in many countries. See, \textit{e.g.}, the references in \cite{ijc13}, \cite{alinea} and the references therein, and\cite{sa,younes,naps,tn,cao,jp,kr,siaap,kim,lis,liu,maa,ieee17,michel17,michel16,siam,portugal,pol,precup17,roumanie,roman0,roman,tun,tapak,ann,lss,exo,buc,nk,xin,yu,zhang,neuro,zhou}, \dots  \ Let us emphasize here \cite{Lafont} and \cite{naps}.

Our communication is organized as follows. Section \ref{mod} describes a simple linear model for the purpose of computer simulations which are analyzed in Section \ref{sim}. Comparisons may be found there with proportional-integral (PI) controllers and with flatness-based control.  Some concluding remarks are presented in Section \ref{conc}. The large literature, that is already available on model-free control, explains why we have not incorporated any summary on this topic and on the related intelligent controllers. Any interested reader in an easy-to-understand introduction is invited to look at \cite{Lafont} and \cite{naps}.

\section{Modeling for computer simulations}\label{mod}
Several authors (see, \textit{e.g.}, \cite{underwood}) emphasized the necessity to simplify a ``full'' mathematical description, which comprises partial differential equations like, of course, the Fourier heat equation. Many publications have been devoted to the derivation of a ``simple'' but ``efficient'' modeling (see, \textit{e.g.}, \cite{Ola_thesis,Ola14,tash,wu}, and the references therein). Like in \cite{Ola16}, a linear model due to \cite{crabb} is used:
{\small \begin{equation}\label{ther_model}
\left\{\begin{array}{cc}
\dot T_{\rm int}=& \frac{Q}{C_a}-\frac{K_c}{C_a}\left(T_{\rm int}-T_{\rm wall}\right)-\frac{K_f}{C_a}\left(T_{\rm int}-T_{\rm ext}\right) \\ \\
\dot T_{\rm wall} =  & \frac{K_c}{C_w}\left(T_{\rm int}-T_{w}\right)-\frac{K_{ext}}{C_a}\left(T_{\rm wall}-T_{\rm ext}\right) 
\end{array}\right.
\end{equation}}

\noindent $T_{\rm int}$, which has to be controlled via the the heat input $Q$, (resp. $T_{\rm ext}$, $T_{\rm wall}$) is the inside (resp. outside, wall) temperature. The other coefficients are suitable physical quantities. Their nominal values are borrowed from  \cite{Radu}: $C_a = 1400$, $C_w = 2200$, $K_c = 1.4$, $K_f = 0.004$, $ K_{ext} = 0.04$.

$T_{\rm ext}$ should be viewed as a \emph{perturbation} in the control-theoretic meaning of this word, \textit{i.e.}, $T_{\rm ext}$ is not regulated.


\section{Comparisons of computer simulations}\label{sim}
\subsection{Intelligent proportional controller}\label{IP}
\begin{figure*}
\begin{center}
\subfigure[Control]{
\resizebox*{8.074cm}{!}{\includegraphics{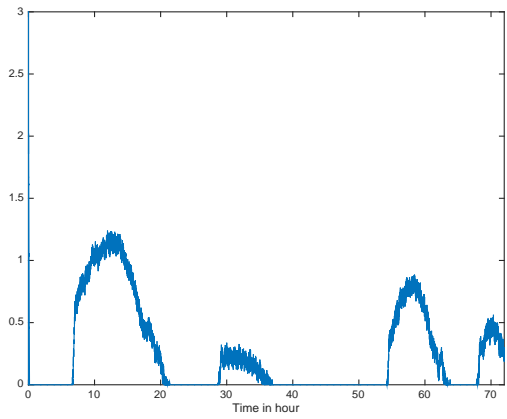}}}%
\subfigure[Reference (- -) and temperature]{
\resizebox*{8.074cm}{!}{\includegraphics{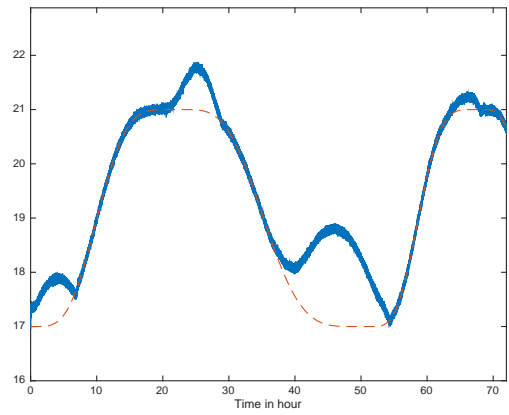}}}%
\caption{iP with heating}
\label{csm}
\end{center}
\end{figure*}

\begin{figure*}
\begin{center}
\subfigure[Control]{
\resizebox*{8.074cm}{!}{\includegraphics{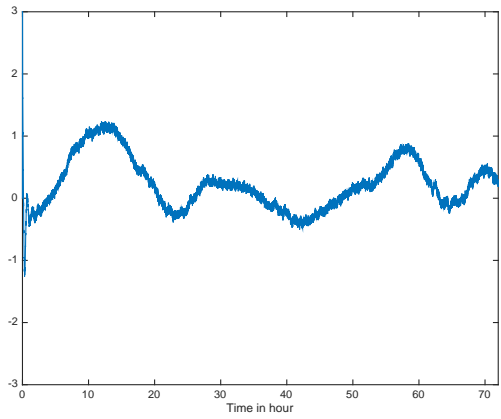}}}%
\subfigure[Reference (- -) and temperature]{
\resizebox*{8.074cm}{!}{\includegraphics{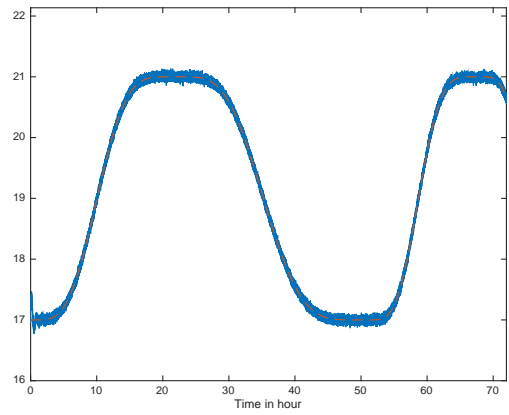}}}%
\caption{iP with heating \underline{and} cooling}
\label{csmn}
\end{center}
\end{figure*}

Start with the most important feedback device in model-free control, \textit{i.e.}, the \emph{intelligent proportional controller} (see \cite{ijc13}, and \cite{naps,Lafont}), or \emph{iP}:
\begin{equation}\label{ip}
u = - \frac{F_{\text{estim}} - \dot{y}^\star - K_P e}{\alpha}
\end{equation}
where $y^\star$ is the reference trajectory, $e = y - y^\star$ is the tracking error, the usual tuning gain $K_P$ and the parameter $\alpha$ are chosen by the practionner, $F_{\text{estim}}$ is an estimate of $F$ in the \emph{ultra-local model} $\dot{y} = F + \alpha u$.
Set in Eq. \eqref{ip} $\alpha = 0.5$, $K_P = - 0.5$. It yields Figure \ref{csm} where the results are excellent with the the exception of two large errors in the tracking: $T_{\rm ext}$ becomes too large and the control variable $Q$ is non-negative. Figure \ref{csmn} shows that those errors disappear if cooling is allowed, \textit{i.e.}, if the sign of $Q$ is arbitrary.
\begin{remark}
Lack of space prevents us from showing the excellent results obtained with values of the coefficients in Eq. \eqref{ther_model} which are quite far from the nominal ones. Moreover, the tuning of our iP does not need to be modified. Parameter identification becomes thus irrelevant.
\end{remark}

\subsection{PI}
\begin{figure*}
\begin{center}
\subfigure[Control]{
\resizebox*{8.074cm}{!}{\includegraphics{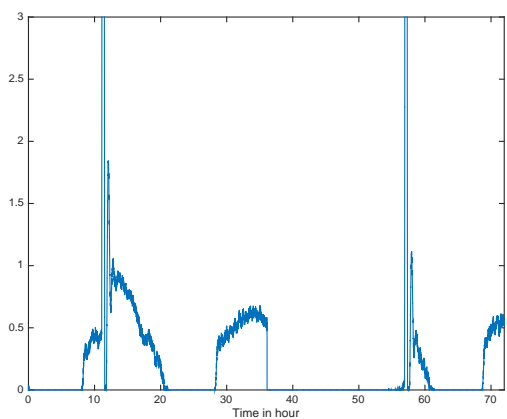}}}%
\subfigure[Reference (- -) and temperature]{
\resizebox*{8.074cm}{!}{\includegraphics{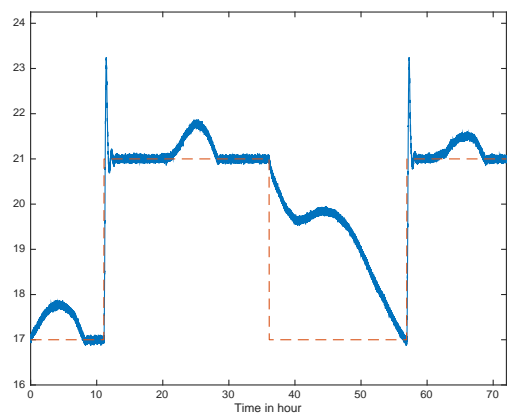}}}%
\caption{PI with steps}%
\label{pistep}
\end{center}
\end{figure*}

\begin{figure*}
\begin{center}
\subfigure[Control]{
\resizebox*{8.074cm}{!}{\includegraphics{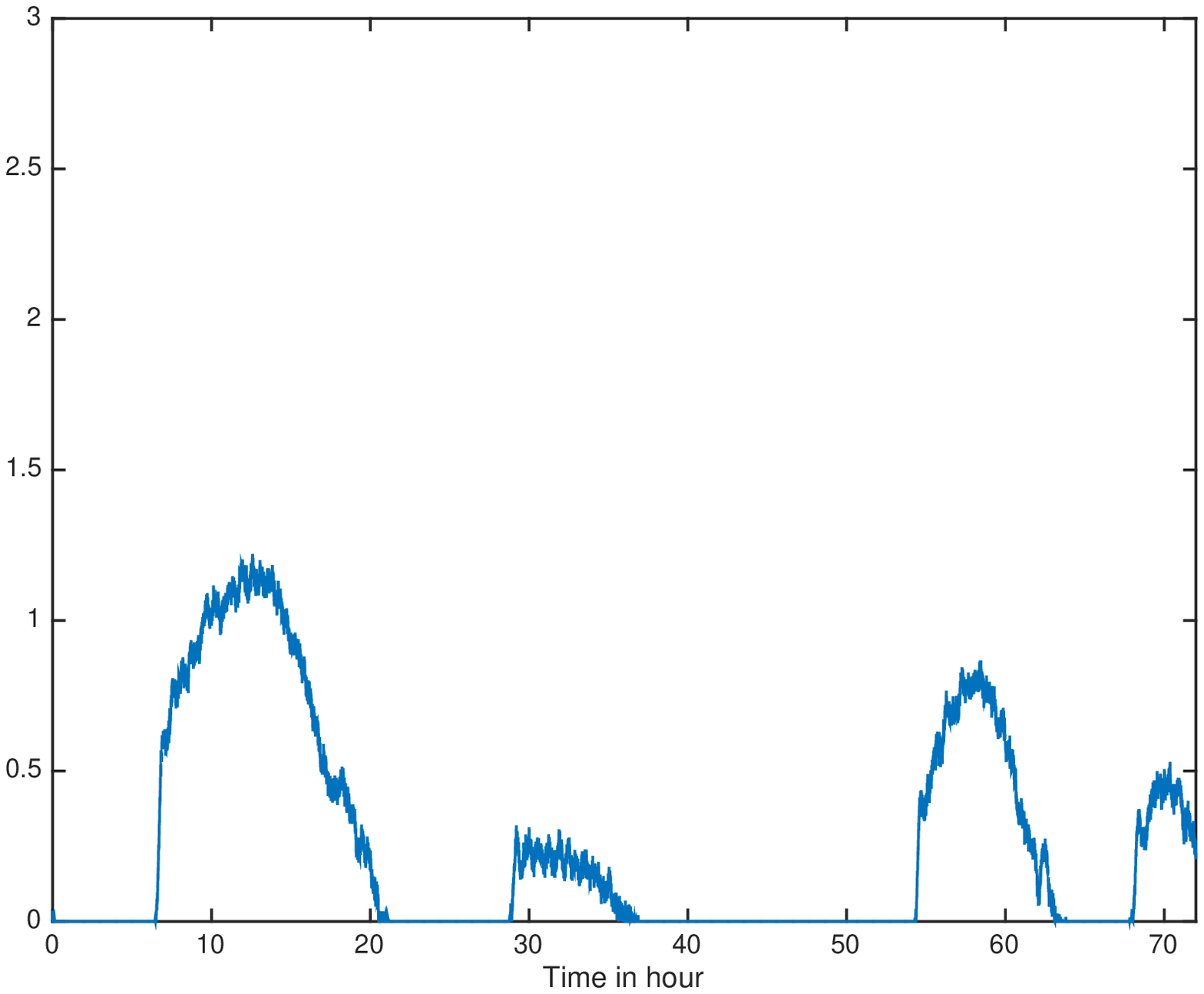}}}%
\subfigure[Reference (- -) and temperature]{
\resizebox*{8.074cm}{!}{\includegraphics{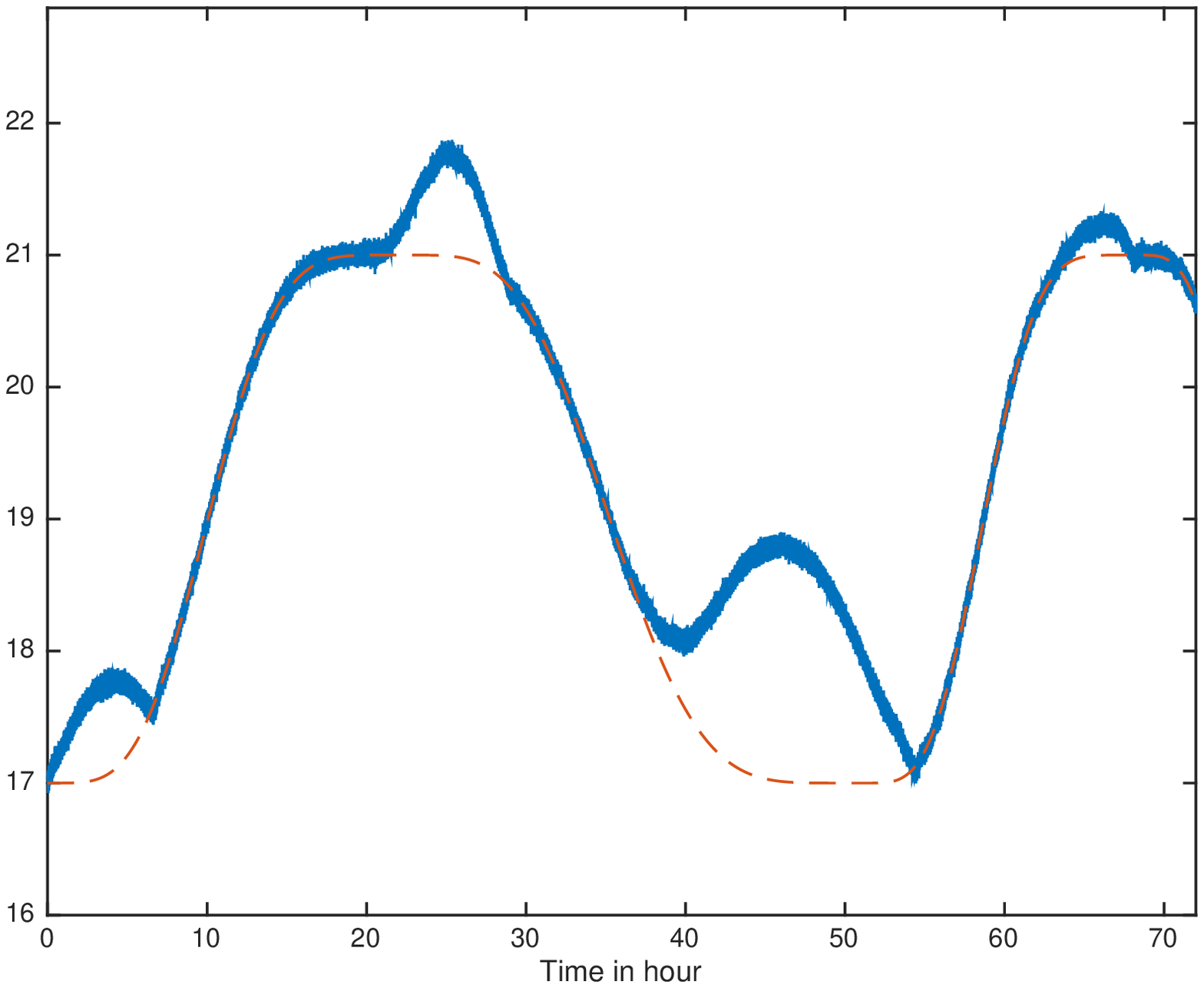}}}%
\caption{PI with a smooth reference trajectory}%
\label{pi}
\end{center}
\end{figure*}

Introduce the classic proportional-integral (PI) controller (see, \textit{e.g.}, \cite{murray}):
\begin{equation}\label{princeton}
Q = k_p e + k_i \int e
\end{equation}
where $e = y - y^\star$ is the tracking error, $k_p$, $k_i$ are the usual tuning gains.
Set $k_p = -0.5$, $k_i = -0.01$. If, like too often in the industrial world, we leave a step between two different setpoints, Figures \ref{pistep} and \ref{csm} show a significant performance deterioration. When this step is replaced by the same smooth reference trajectory used in Section \ref{IP}, Figures \ref{pi} and \ref{csm} become almost identical. This result is not surprising. It is known \cite{ijc13} that iPs and PIs are equivalent in some sense.

\begin{remark}
In order to avoid steps we could have also employed \emph{setpoint ramping} (see, \textit{e.g.}, \cite{ramp}).
\end{remark}
\subsection{Flatness-based predictive control}\label{flat}
\subsubsection{Generalities}
Flatness-based control, which was introduced more than 25 years ago \cite{ijc95}, certainly is one of the very few academic advances which have become popular in the industrial world (see also, \textit{e.g.}, \cite{murray,ijc00,levine,riga,sira-flat,tarbes,zeitz}). It might be viewed as another way of doing predictive control, but without any optimization technique. 
From this standpoint, the first Eq. in \eqref{ther_model} yields 
$$
Q^{\star} = {C_a}\left(\dot T_{\rm int}^{\star}+(\frac{K_c}{C_a}+\frac{K_f}{C_a})T_{\rm int}^{\star}\right)
$$
where $T_{\rm int}^{\star}$ is the reference trajectory for the internal temperature and $Q^{\star}$ the corresponding open-loop control. The closed-loop control reads $Q = Q^{\star} + C(e)$, where $C(e)$ is some regulator.
\subsubsection{P}
\begin{figure*}
\begin{center}
\subfigure[Control]{
\resizebox*{8.074cm}{!}{\includegraphics{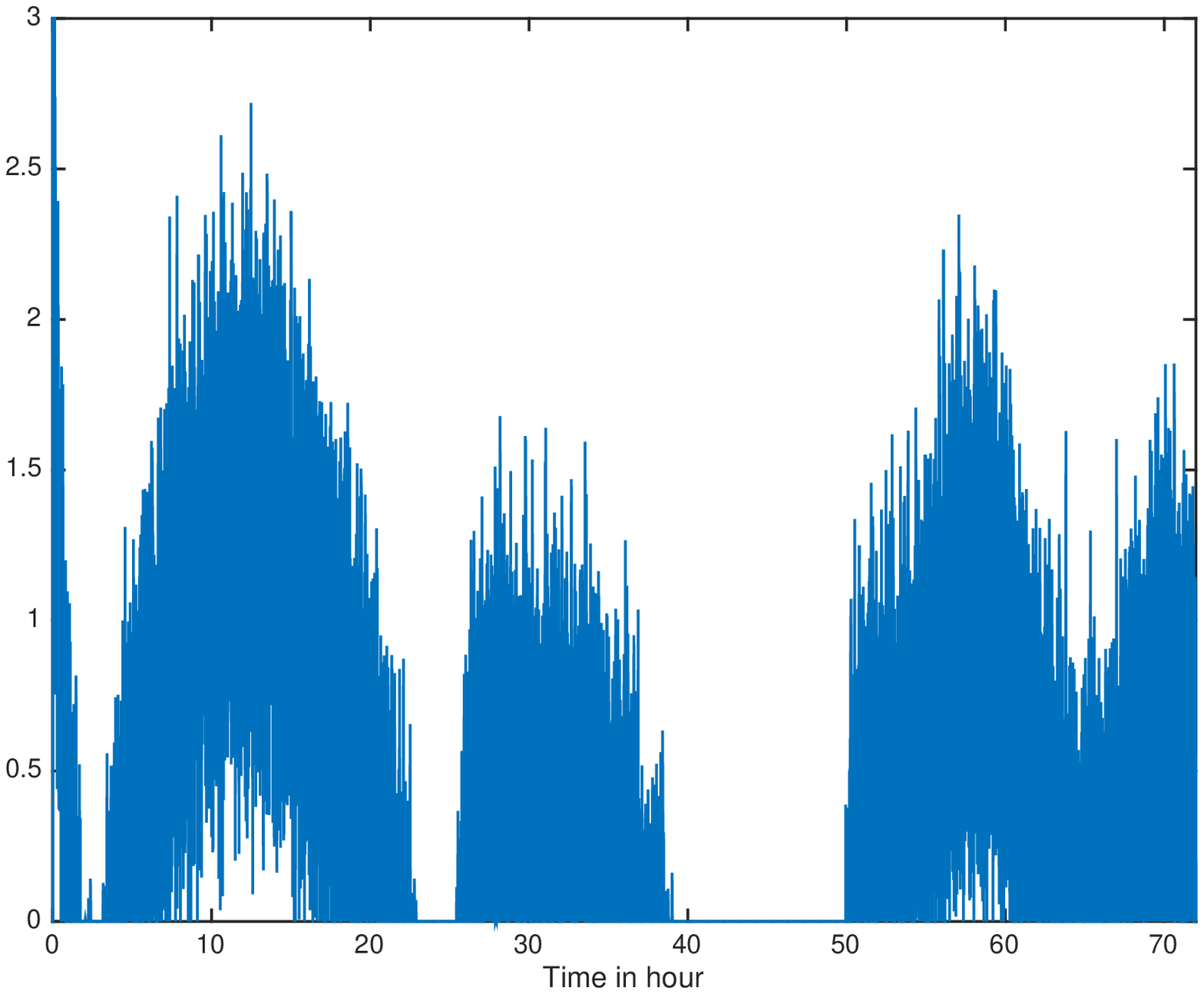}}}%
\subfigure[Reference (- -) and temperature]{
\resizebox*{8.074cm}{!}{\includegraphics{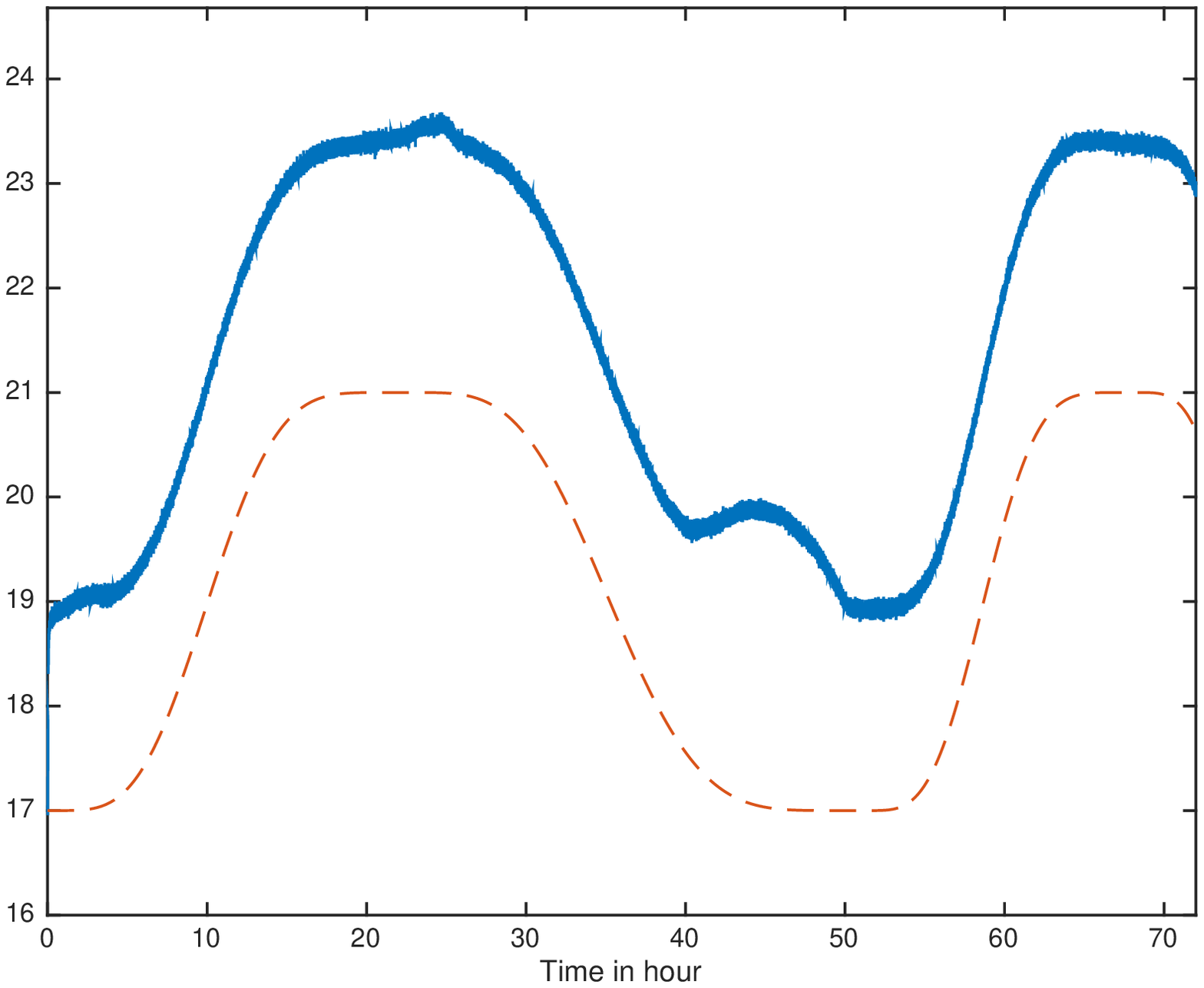}}}%
\caption{Flatness and P}%
\label{fp}
\end{center}
\end{figure*}

$C(e)$ is a static state-feedback, \textit{i.e.}, nothing else than a trivial P controller. Determine it by imposing a pole equal to $0.01$. The results in Figure \ref{fp} are poor. The perturbation represented by the external temperature cannot be rejected if it is not measured.
\begin{remark}
Let us add however that such a perturbation might be measured and therefore rejected via some appropriate tools (see, \textit{e.g.}, \cite{easy}).\footnote{Such a control design is often called \emph{Active Disturbance Rejection Control}, or \emph{ADRC} (see, \textit{e.g.}, \cite{adrc} and the references therein).}
\end{remark}
\subsubsection{PI} 
\begin{figure*}
\begin{center}
\subfigure[Control]{
\resizebox*{8.074cm}{!}{\includegraphics{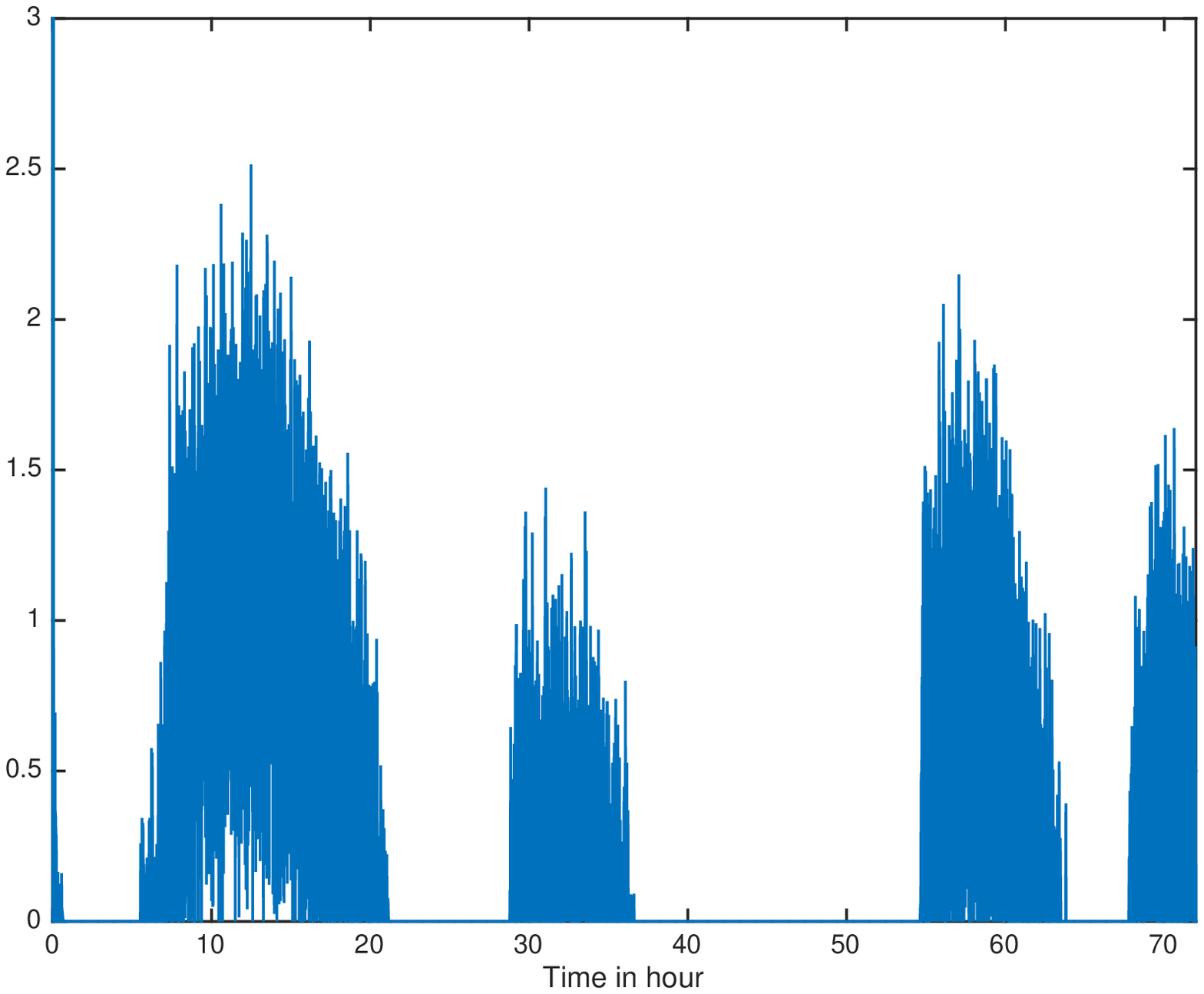}}}%
\subfigure[Reference (- -) and temperature]{
\resizebox*{8.074cm}{!}{\includegraphics{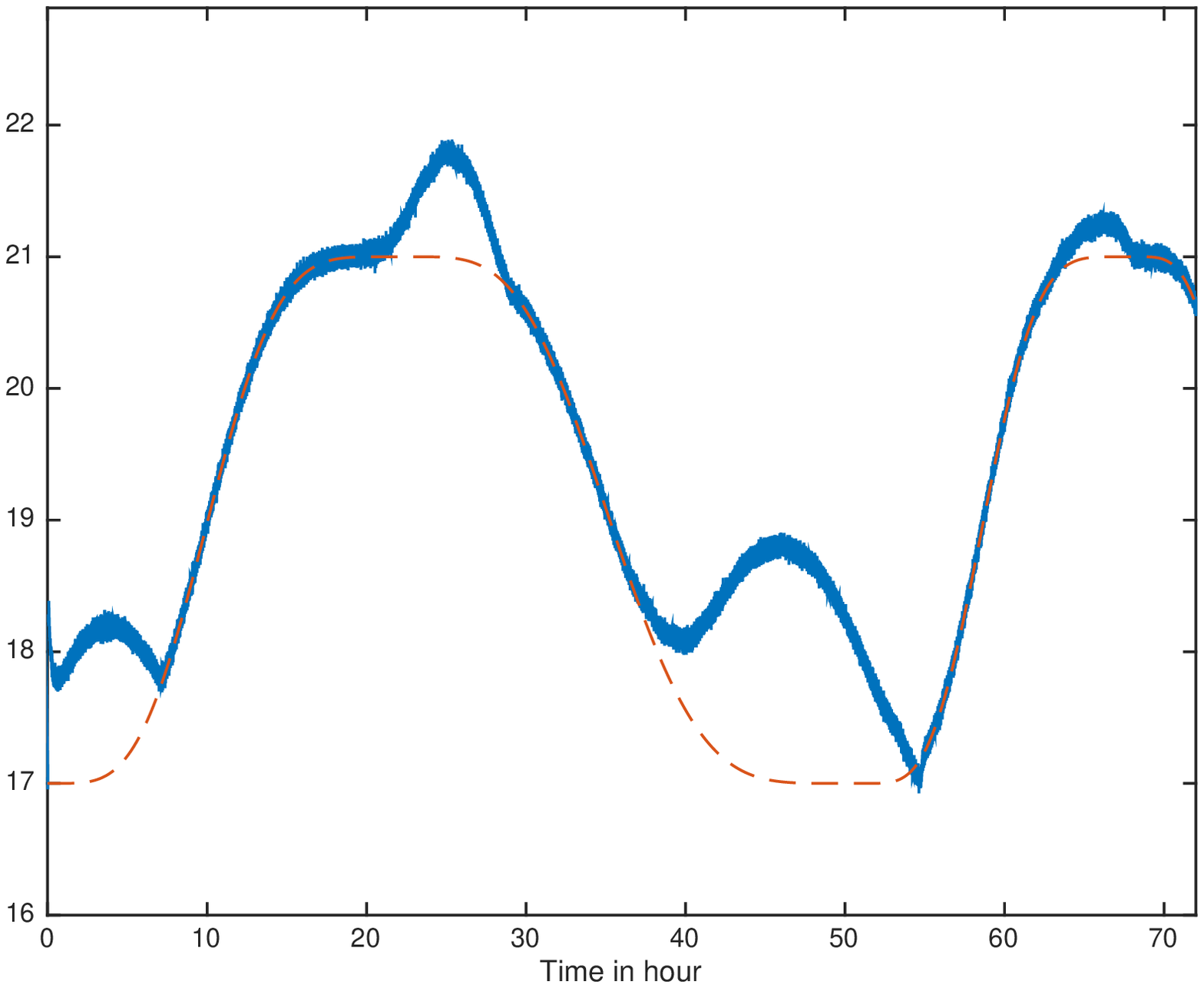}}}%
\caption{Flatness and PI}%
\label{fpi}
\end{center}
\end{figure*}

\begin{figure*}
\begin{center}
\subfigure[Control]{
\resizebox*{8.074cm}{!}{\includegraphics{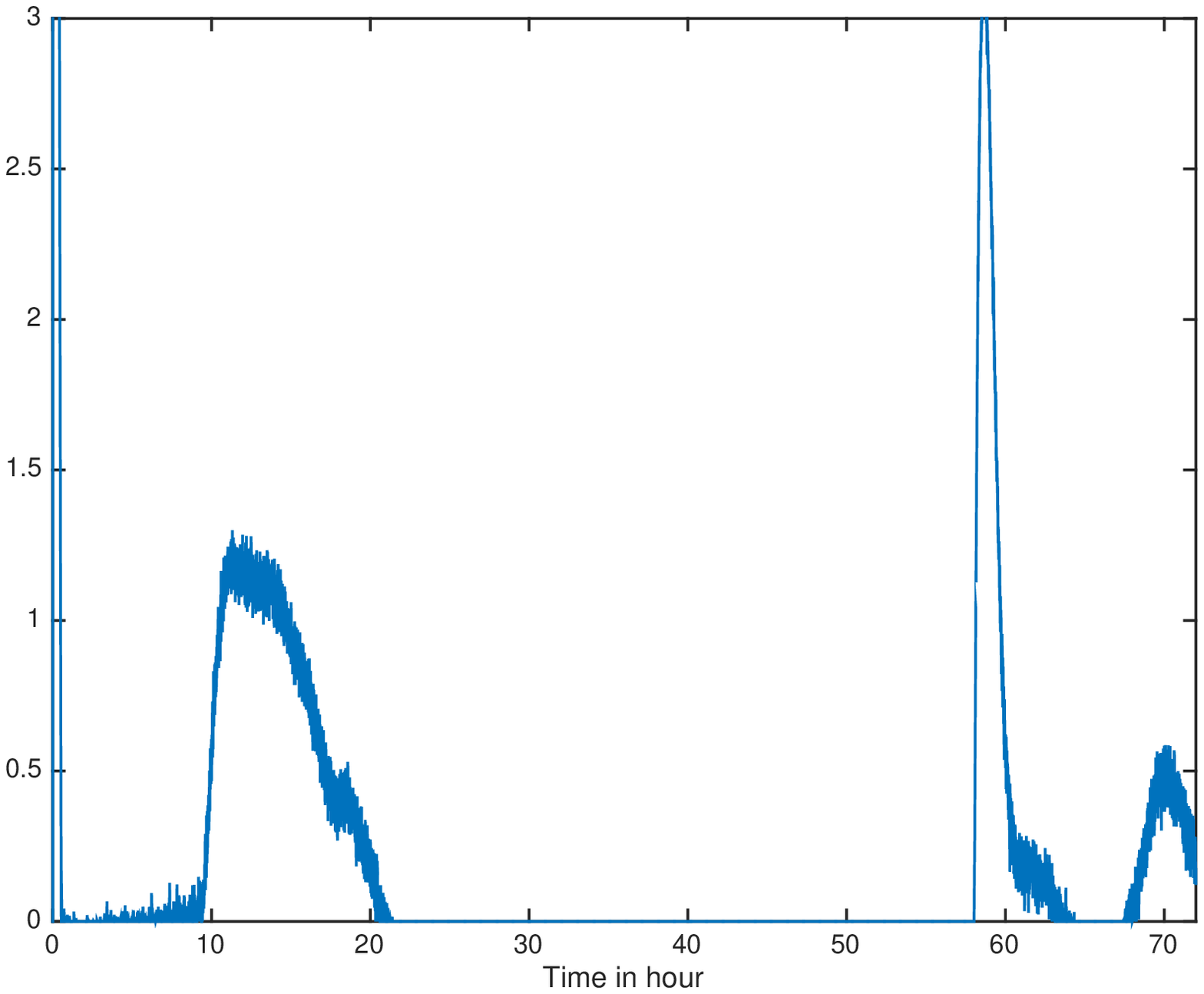}}}%
\subfigure[Reference (- -) and temperature]{
\resizebox*{8.074cm}{!}{\includegraphics{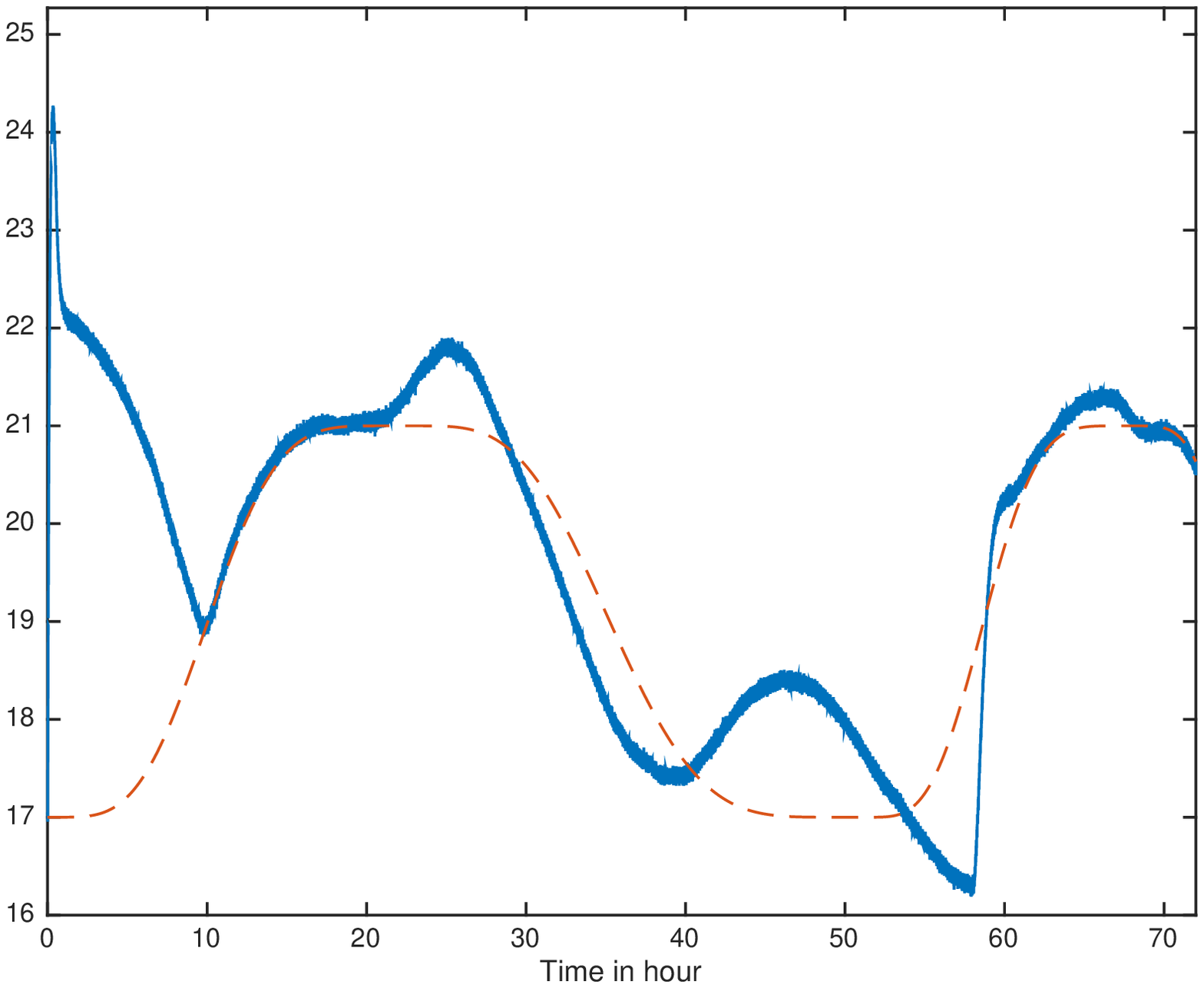}}}%
\caption{Flatness and PI with ``small'' poles}%
\label{fpip}
\end{center}
\end{figure*}

$C(e)$ is now a PI \eqref{princeton}. The results in Figure \ref{fpi}, where the gains are tuned such that there is a double pole equal to $-0.005$, are rather good. Note however that the control input fluctuates too quickly. With a double pole equal to $-0.001$, the comparison of performances displayed in Figure \ref{fpip} and Figure \ref{csm} show a poor robustness with respect to noises.
\begin{remark}
Section \ref{flat} confirms that a model-based standpoint might be cumbersome and, therefore, inappropriate.
\end{remark}





\section{Conclusion}\label{conc}
Satisfactory computer simulations were presented via model-free control and with a great ease. They will soon be tested in a real building and hopefully confirmed. The excellent results which were recently obtained with respect to an experimental greenhouse  
\cite{Lafont} should make us optimistic.



%

\end{document}